\documentstyle[12pt]{article}
\begin{document}

\title{New theory of periodical systems by finite
interval inverse problem.}

\author{B. N. Zakhariev and V. M. Chabanov\\ Laboratory of
theoretical physics, JINR, \\ Dubna, 141980, Russia; e-mail:
zakharev@thsun1.jinr.ru } \date{April 2, 2003} \maketitle

\section{Abstract}
We show that the mechanism of gap formation  has a {\it resonance}
nature. The  special real fundamental solutions were discovered
which `paradoxically' have knot distribution with a period
coinciding with that of potential at all energies of the whole
lacuna interval. In terms of these solutions resonance gap
appearance gets the most direct explanation: ever repeating hits
by the potential result in exponential increase (decrease) of the
wave amplitudes in the forbidden zones. The analogous alternating
hits from opposite sides are responsible for the wave beatings in
allowed zones. The inversion technique gives rise to zone control
algorithms -- shifting chosen boundaries of spectral bands,
changing degree of zone forbiddenness. All this cannot be achieved
by the previous Bloch theory.

\section{Introduction}

It is well known  that periodical potential perturbation splits
the continuous spectrum creating bands of allowed zones separated
by gaps of forbidden zones. The explanation of this statement by
Bloch-Floquet formalism  has seemed to be satisfactory since the
appearance of this theory 70 years ago. However, the intrinsic
mechanism of zone formation by a certain periodic potential
remained incompletely understandable: Bloch theory as a black box
gives out only the spectral zones without qualitative
explanations. Additional insight into the theory became possible
due to inverse problem and SUSY QM formalisms that opened new
horizons in quantum mechanics by their exactly solvable models,
see  our new book : "Submissive quantum mechanics" about the
turnover from direct to inverse problem \cite{obed} (it can be
found  in internet http://thsun1.jinr.ru/$\sim$ zakharev/).

We have found the following fact: there is some kind of
deep-hidden 'resonance' which tears the spectrum. Namely, inside
the gaps the period of the potential coincides with that of very
special (we will call them `smart', i.e. `intellectual')
solutions. Being periodically positioned, knots of these smart
solutions are not the same, but interlaced with each other.
 Of these two linearly independent solutions, being a
manifold of zero measure, the whole continuum of solutions with
'invisible resonance' can be constructed. They will have quite
irregular disposition of knots at each energy value in the
forbidden zone. Despite this 'irregularity camouflage' due to a
resonant nature of constituents (coherent hits of perturbation in
the successive periods), `smart' solution oscillation amplitudes
are forced to exponentially increase (or decrease)in opposite
directions. This is characteristic to non-physical (forbidden)
behaviour. Hence, arbitrary solutions tend asymptotically (as $x
\to \pm \infty$) to pure smart solutions. On the one side, the
discovered resonance mechanism seems to be very natural. On the
other side, the property of smart solutions having the same period
of knots (oscillations) on the finite energy interval contradicts
the widely spread notion that the oscillation frequency increases
with the energy, for instance, the initial free solution  has a
specified period at only one fixed energy point. In periodical
potentials there are almost no periodical solutions except the
zone boundaries. Bloch waves are products of functions with
different periods. In forbidden zone nobody suspected of any
periodicity. The unexpected periodical positioning of knots in
rare smart solutions reveals almost evidently the resonance cause.

The above mentioned can be explained in the following way.  {\it
Different solutions perceive the same potential as different
effective potentials}: more or less repulsive (attractive). It is
due to different sensitivity of bumps and knots of (real) solution
to the local values of the potential. So, for instance, if bumps
of the smart solutions are at potential barrier positions there
will prevail a repulsion, i.e., the effective kinetic energy will
be weakened, which widens the distance between knots. This allows
{\it energy change complete compensation} needed to keep knot
period constant all over the energy interval of the entire
forbidden zone. So the inter-knot distance of smart solutions
remains unchanged when shifting simultaneously the energy over the
$E$-axis and the wave knot lattice over the $x$-axis.

It is worth to mention that the choice of purely real solutions
was one of the key moments simplifying the investigations of this paper.

Splitting the unperturbed continuous spectrum into the band one by
{\it creating forbidden zones is essentially quantum effect}.  So
we shall return later once more  to clarify this basic features of
wave behavior in periodical structures which is not at all
'non-physical' aspect of the theory.  We shall also consider below
somewhat less wonderful solutions in the allowed zones.

\section{Inverse problem and SUSY QM  spectral control}

The complete sets of explicit inverse problem solutions on finite
interval served us as a hint in  constructing  solutions to
periodic potentials. Moreover, they give rise to a significantly
broader class of exactly solvable models for periodic systems
(infinite zone models).

The key point of the inverse problem is that spectral parameters
(energy levels, normalizing constants, etc.) enter in the
formalism as input data and corresponding exactly solvable models
are generated by varying only a certain spectral parameter or a
finite number of them (while living the others unaltered). Since
spectral parameters form a complete set and completely determine
properties of quantum systems, exact models corresponding to all
possible variations of spectral parameters give rise embrace the
whole manifold one- and multi-channel systems. Thus, there appears
a possibility of changing at wish quantum objects by variation of
these parameters (as if using control 'levers') and examine
quantum systems in different thinkable situations. By the way,
instead of about ten direct problem exactly solvable models which
currently serve as a basis of the contemporary education one
should also be guided by an infinite (!) number, even {\it
complete} sets of such models from the inverse problem.

The laws of  structure transformations and wave motions revealed
by computer visualization of these models were reformulated into
unexpectedly simple universal rules. The elementary  "bricks and
blocks" were discovered of which it is possible in principle to
construct objects with any given properties \cite{CZS}.

Let us shortly formulate these rules \cite{obed,Les} because they
will serve as suggesting considerations in constructing remarkable
solutions with periodically spaced knots. Each real wave consists
of bumps and knots. We now discuss two algorithms of
transformation of {\it one} wave bump . They are easily
demonstrated by the example of eigenvalue problem on a finite
interval with zero boundary conditions (namely, for the ground
state of the infinite rectangular well). First rule: {\it To shift
one energy level up (down)} over the energy scale we must
introduce a potential barrier (well) near the extremum of the bump
where the wave is most sensitive to the potential ($ V$), see
Fig.1. Pay attention to the important immobility of the ground
state knots that promoted us in revealing resonance ('smart')
solutions. In addition, we need compensating wells (barriers) in
the vicinity of knots if we want to keep all other energy levels
at their previous places. So {\it the different states see the
same potential as different effective potentials}.  The {\it
second rule}:  for {\it shifting the spatial localization of a
wave bump} over the finite interval to the right (left) we need
potential perturbation as a barrier-well (well-barrier) block.
{\it These rules are applicable to wave bumps of each (bound or
scattering) state and for any potential} \cite{Les}, see Fig.1.,
Fig.2.

 As a result, we can transform any wave with arbitrary number of bumps. So
we acquire the vision of the intrinsic logic of structure and
behavior peculiarities of a great amount of thinkable systems,
including real ones.

\section{Waves in periodical potentials}

Periodical perturbation 'stretches one spectral point' of initial
free motion into continuous energy interval of forbidden zone with
zero spectral density; in  allowed zones the density is compressed
\cite{Levit}.

 Switching on the perturbation violates generally the periodicity
(equidistancy) of knots of the real solutions. There are continuum
such solutions at any energy value and all solutions, unlike those
for  free motion, are quite different (they do not turn into one
another by shifting $x$-variable).

Consider a free one-dimensional motion of waves over the x-axis.
The Schr\"odinger equation has two linearly independent partial
solutions. It suffice to fix the position of any  knot to
determine the solution up to an inessential norming factor. For
zero potential there is a continuous spectrum $E>0$. A typical
solution is $\sin(kx+\alpha )$ with the phase $\alpha $.

We shall choose, e.g., the periodic potential with the
Kronig-Penney form, see Fig.3.

 If the wave bumps (their extrema) are
positioned  above the  middles of potential wells (see Fig.4C) and
the knots inside the potential barriers, the wave 'must have'
higher average kinetic energy and, hence, frequency of
oscillations. Thus, we attained the same frequency as was in free
motion but at a lower energy value $E_{<}$. This corresponds to
pushing downward to the energy $E_{<}$ the ground state of the
auxiliary problem of the infinite rectangular potential of width
equal to the period (see Fig.1) with the potential wall positions
coinciding with the knots. Remind that such a shift of levels is
performed without moving the knots at the ends of the interval
\cite{CZS}. On the contrary, shifting the bump extrema of
$\sin(kx+\alpha )$ to the middles of the barriers (see Fig.4A)
would decrease in the average kinetic energy if we keep the energy
$E=k^2$ unchanged. Consequently, one has to lift energy up to
$E_{>}$ for the inter-knot distance to remain unaltered.
 The symmetry of potential between the knots according to
the middle point between the knots in both the cases conserves the
symmetry of the corresponding solutions on the periods. This
allows their smooth periodic continuation. The solutions
considered correspond to the upper and lower boundaries of the
forbidden zone which is created by the potential perturbation.

This is confirmed by the consideration of other solutions with the
energies between $E_{>},\,E_{<}$.  Take the solution with the
knots at the boundaries of the potential wells and barriers (see
Fig.4.B,D).  In this case, under each bump is the block
"well-barrier" or "barrier-well" which slightly change the energy
of the state which follows from the requirement of keeping
inter-knot distance unaltered. Besides this there is  strong
violation of the symmetry of the bump, in particular, its
derivatives at the knots (see discussion below). This leads to
exponential growth of the solution oscillation amplitude as $x \to
\pm \infty$. It follows from smooth connection of solutions on the
neighbor periods. All other dispositions of the knots with respect
to wells and barriers of the potential will lead to combinations
of shifts over the energy scale and violations of symmetry of the
initial bumps. So the whole spectral lacuna can be continuously
filled by these solutions with gradual change in the 'degree of
forbiddeness' of the zone from zero at one boundary to the maximum
value somewhere near its middle and again to zero at the other
boundary. This happens because of the increase-decrease of
symmetry violation of the perturbing potential between the knots
(pay special attention to the gradual shift of state over E and to
the movement of bumps localization over x). Here the
mathematically exact proof can be obtained without formulae
(explanations 'on fingers' based, however,  on the previous rules
of quantum intuition).
 Arbitrary solution  in forbidden zone has different asymptotically
periodical  displacement  of knots because only increasing
component of the corresponding linear combination of smart
solutions survives at asymptotic distances (solution tends to one
or another smart solution as $x \to \pm \infty$).

Above $E_{>}$ inside the allowed zone the frequency of knots
begins to increase with $E$ until the number of knots on one
period increases by one.  The solutions there have alternating
increases and decreases of oscillation amplitudes. These beatings
in allowed zones can be explained in the following way. There are
no analogies of 'smart' solutions with equidistant knots except
the zone boundaries  where there is one periodic solution among
the continuous set with disordered knots. With increase of energy
the (average) density of 'disordered' knots also increases, as in
free motion. There is difference in size of the potential period
and the distances between solution knots. On the one side, it
leads  to oscillations of these knot positions with respect to
potential periods along the $x$ axis. On the other side, there is
alternation of regimes of increase-decrease of oscillation
amplitudes. The asymmetry of potential perturbations between the
neighbor knots and inequality of solution derivatives there result
in increasing and then decreasing (and so on) of swinging of the
solutions, i.e., lead to beatings.  In other words, at some
moment, there may be more effective  attraction, say, to the left
and repulsion from the right between the knots of the solution.
This will make the solution oscillation amplitude increase to the
left. Later, due to shifting oscillations of the solution relative
the potential periods, the situation changes to the opposite one:
there will be excess of repulsion from the left and attraction
from the right. The wave length of these modulated oscillations
(beatings) will decrease from infinite value on the lower boundary
of the allowed zone to a finite value of about the potential
period size at the upper boundary.

At the energy zone boundaries, in addition to periodic solution,
there is a solution linearly increasing in both directions. This
reminds the situation with the solutions in vicinity of eigenstate
with an energy $E_{\nu }$ in a finite depth potential well. Above
and below $E_{\nu }$ there are two independent solutions
exponentially increasing (decreasing) to the right (left) sides
and vice versa. Meanwhile  at $E_{\nu }$ there are two solutions
increasing and decreasing (bound state proper) in both directions.

Let us  discuss once again the mechanism of zone formation by
using somewhat different scheme. Let us at first consider an
auxiliary problem of finding solutions and spectrum of the
Hamiltonian with zero boundary conditions at the edges of the
finite interval (period) $[\varepsilon, a + \varepsilon]$, for any
$0 < \varepsilon < a$ and the potential
$V^{\varepsilon}(x)=V_{per}(x)$, e.g., Kronig-Penney one, for $x$
belonging to this interval. The corresponding spectrum is discrete
set $\{E_{n}^{\varepsilon }\}_{n=0}^{\infty}$. If we continue
periodically (with the period $a$) this interval we will, of
course, restore the periodical potential and, moreover, find
particular solutions for $V_{per}(x)$ at the energies
$E_{n}^{\varepsilon }$. In fact, the sought solutions may be
obtained from the auxiliary problem eigenfunctions $\Psi
^{\varepsilon} (x,E_{n}^{\varepsilon})$. The wave continuity is
guaranteed by vanishing of the eigenfunctions at the matching
points $n a + \varepsilon$ (points of contact of adjacent
intervals). The smoothness of wave matching can be attained by
multiplying neighbor  eigenfunction by a factor being equal
(modulo) to the ratio of the derivatives at the right and the left
boundaries. If this factor is  equal modulo to $1$ the energy
$E_{n}^{\varepsilon}$ will be at one of the zone boundaries. This
case corresponds to such position of the interval $[\varepsilon, a
+ \varepsilon]$ that $V^{\varepsilon}(x)$ is a symmetric (with
respect to the interval center) potential with a rectangular
barrier (or well) in the middle, see Fig 4 A,C. When once the
potential symmetry is violated (by altering $\varepsilon $), $\Psi
^{\varepsilon} (x,E_{n}^{\varepsilon})$ loses its previous
symmetry, which results in either increasing or decreasing
solution on the whole axis. Hence, the  energy
$E_{n}^{\varepsilon}$ turns out to be inside the forbidden zone.
Simultaneously, the $E_{n}^{\varepsilon}$ is shifted downward
(upward) since the potential $V^{\varepsilon}(x)$ becomes
effectively less repulsive (attractive) as a result of shifting
barrier (well) position on the interval. Thus, by varying the
parameter $\varepsilon $ (as if scanning the potential by the
interval $[\varepsilon, a + \varepsilon]$), the energy
$E_{n}^{\varepsilon}$ runs the entire nth spectral gap.

\section{Shifts of zone boundaries}

It appears that our ability to shift isolated bound state energy
levels (over the $E$-scale) in the infinite potential well on a
finite interval allows us to move some chosen upper (lower)
boundaries of the spectral zones (bands) of periodic structures
keeping infinite number of other upper (lower) boundaries
unperturbed.

For definiteness, we will consider shifting up and down the upper
boundary of only the second allowed zone for the initial Dirac
comb of periodic $\delta $-barriers. For this purpose, we will use
the SUSY QM formulas for only chosen energy level shift on the
finite interval $[0,\pi]$ with potential $V_{0}(x)$ \cite{Trub}.
Let $\psi_{0}(x,E_{n})$ be corresponding eigenfunction at the
energy $E_{n}$. We assume ${\bar \psi}_{0}(x,E_{n}+t)$ to be a
non-physical auxiliary solution in the initial potential at
shifted energy $E_{n}+t$ with the symmetry being opposite to that
of the $\psi_{0}(x,E_{n})$.  With these solutions, we can
construct the Wronskian
\begin{eqnarray} \theta(x)
\enskip = \enskip \psi_{0}'(x, E_{n}) {\bar \psi}_{0}(x,E_{n}+t)
\nonumber
\\ - \psi_{0}(x, E_{n}) {\bar \psi}_{0}'(x,E_{n}+t).  \end{eqnarray}
Shifting energy level $E_{n} \to E_{n}+t$ is performed by special
potential transformations. The corresponding final expressions for
the transformed potential and solutions read \begin{eqnarray} V(x)
\enskip = \enskip \nonumber
\\ V_{0}(x) - 2 \, t \, \frac{d}{dx} \{ \psi_{0}(x, E_{n}) {\bar
\psi}_{0}(x,E_{n}+t) \theta(x)^{-1} \}; \label{v2} \end{eqnarray}
\begin{eqnarray} \psi(x,E) = \psi_{0}(x,E) - \nonumber \\ t {\bar
\psi}_{0}(x,E_{n}+t) \theta(x)^{-1} \int_{0}^{x} \psi_{0}(y,
E_{n}) \psi_{0}(y,E)dy; \label{psi2} \end{eqnarray}
\begin{eqnarray} \psi(x,E_{n}+t) \enskip = \enskip \psi_{0}(x,
E_{n}) \theta (x)^{-1}. \label{psit} \end{eqnarray} The potential
(\ref{v2}) can be periodically continued:  $V^{per}(x+l\pi)=V(x),
\enskip l=0; \pm 1; \pm 2;...$ for $x$
 belonging to $[0,\pi]$. Such a scheme allows a spectral zone control.
Let us consider this potential with an incorporated Dirac comb
(equally spaced $\delta $-potential barriers or wells
$\sum_{m=-\infty}^{\infty} v \delta (x-m\pi)$), i.e., the
following periodic potential: $\sum_{m=-\infty}^{\infty} v \delta
(x-m\pi) + V^{per}(x)$. The corresponding results are shown in
Fig.5.

In the special case of {\it mergence of allowed zones} there is
local compensation of attraction and repulsion, at the energy
value $E_{m}$ and $\Delta E=1$ :
 as in the case of a free motion, any choice of boundary condition
 (position of some knot) gives solutions on periods with symmetric
derivatives at the ends of the interval (their invariance under
arbitrary translations),  which provides wave behavior without
exponential growth. Any small change of energy leads to violation
of this situation.  A like mergence of the first and the second
allowed zones see Fig.6. (something above $\Delta E=-2$  and below
$\Delta E=3$).

Exactly solvable models allow one to demonstrate the control of
`forbiddenness' at any chosen energy point by changing the
derivative of finite interval  solution at the left boundary, the
spectral weight vector $c$ \cite{IP,Ann}. Below are given the
formulas for the potential and wave functions transformations
corresponding to variations of the spectral weight factor $c_{n}$
of a chosen bound state at the energy $E_{n}$.  Let
$\stackrel{\circ}{\psi }_{m}(x)$ be the eigenfunction of the
initial Hamiltonian corresponding to $m$th energy eigenvalue,
$\stackrel{\circ}{c}_{m}$'s are the initial spectral weight
factors.  Then changing $\stackrel{\circ}{c}_{m} \to c_{m}$ gives
\begin{eqnarray}
 \psi_{n}(x)=\stackrel{\circ}{\psi }_{n}(x)+
\frac{(1-c_{m} ^{2}/\stackrel{\circ}{c}_{m}\!^{2})
\stackrel{\circ}{\psi }_{m}(x)}
{1-(1-c^{2}_{m}/\stackrel{\circ}{c}_{m}\!^{2}) \int
\limits_{0}^{x}\stackrel{\circ}{\psi }_{m}\!^{2}(y)dy} \nonumber \\
\times \int \limits_0^x \stackrel{\circ}{\psi
}_{m}(y)\stackrel{\circ}{\psi }_{n}(y)dy, \label{psi}
\end{eqnarray}
\begin{eqnarray}
V(x)= \stackrel{\circ}{V}(x) \nonumber \\
+ 2\frac{d}{dx}\frac{(1-c_{m}^{2}/\stackrel{\circ}{c}_{m}\!^{2})
\stackrel{\circ}{\psi }_{m}\!^{2}(x)}
{1-(1-c_{m}^{2}/\stackrel{\circ}{c}_{m}\!^{2}) \int \limits_0^x
\stackrel{\circ}{\psi }_{m}\!^{2}(y)dy}. \label{V}
\end{eqnarray}
Again, we periodically continue these results. Changing $c$ leads
to previous symmetry violation for the initial wave function
$\stackrel{\circ}{\psi }_{n}(x)=\sqrt(2/\pi)\sin(nx)$ (we put
$\stackrel{\circ}{V}(x)=0$ and the interval to be $[0, \pi]$). As
a result, the whole axis wave function, `sewed' from $
\psi_{n}(x)$, will asymptotically diverge. The factor $c$  just
represents the index of divergence rate, `forbiddenness'. See
results on the corresponding control in Fig.7, \cite{obed} (in
particular, we can thus tear the continuous spectrum at energy
point $E_{n}$).

Instructive are some simple examples of shifting only  the upper
or only lower boundaries of zones. They can be achieved by
introducing $\delta $-peaks or $\delta $-wells at the middle of
the periodical potential barriers or wells. Take into account that
the $\delta $-potentials do not influence the solution if they are
coincident with the knots, but push the energy of state maximally
up or down, being disposed at the maximum modulo of the wave
bumps.

Analogous reasoning is applicable to other forbidden zones. We
only need to use the initial free sine-like solutions at higher
energies corresponding to knot frequency twice (and more times) in
comparison with the periodic oscillations of the potential.

We can generalize the theory for arbitrary periodic potential for
which it is impossible to choose the period with symmetrical
potential shape (with respect to the center of period).
 In spite of asymmetry in this case there are such positions of solution knots
relative to potential at specific energies $E_{<,>}$ that the
derivatives at knots become equal modulo. So we get  the
boundaries of zones. See Fig.8. as an example potential with peak
of barriers and bottoms of wells at the same points. In this case
$\Psi ; \Psi_{1}; \Psi_{2}; \Psi_{3}$ correspond to spatial shifts
to the right, left and E-shifts up and down.

It is worth to mention the possibility to create localized bound
states in periodic structures in analogy with transformation of
free motion (soliton-like potentials which are transparent for
Bloch waves \cite{Sams1,Samson}.

Everything we have learned about the periodic structures (smart
solutions in forbidden zones, beatings in allowed zones) is
applicable to the finite interval potentials created from
truncated periodic potentials (i.e., becoming periodic when
continuing over the entire $x$-axis), in particular, to resonance
tunneling through finite number of periodic potential barriers
where multiplets of resonances correspond to allowed zones and
gaps between them to forbidden zones.

It is worth to mention that we have   found examples for complex
periodic potentials \cite{Nongam}) which do not create lacunas.

\section{Some multichannel models.}

We hope to generalize our theory to systems of coupled
Shr\"odinger equations. As the first steps in this direction can
be considered the following models illustrating some important
features of complex systems.

1. Let $V_{11}(x)=V_{22}(x)$ (with equal periods) and the
thresholds of both the channels are equal $\varepsilon_{1}=
\varepsilon_{2}=0 $. For uncoupled (independent) channels the
one-channel theory is obviously generalized.  There are separate
spectral bands in both the channel spectral branches.  Now let us
switch on some interchannel coupling $V_{12}(x)=V_{21}(x)$. So,
the interaction matrix is symmetrical with respect to channels. If
we choose  boundary conditions for both the channels equal, the
partial channel functions will be equal $\Psi_{1}(x)=\Psi_{2}(x)$.
This allows one to make the substitution $V_{12}(x) \Psi_{2}(x)
\rightarrow V_{12}(x) \Psi_{1}(x)$ in the equation for the first
channel and analogous transformation in the second channel. This
results in their effective separation (coupling disappearance :
'diagonalization' of the interaction matrix) with new effective
interaction matrix elements $V_{11 eff}=V_{11}(x)+V_{12}(x)$;
$V_{22 eff}=V_{22}(x)+V_{21}(x)$.  So,
 spectral bands of these separate channels will coincide with one another
and for the  whole system.  If we choose different signs for
partial waves (determining corresponding boundary conditions) then
in the new effective diagonal interaction matrix elements there
will be "-" instead of "+" $V_{11 eff}=V_{11}(x)-V_{12}(x)$;
$V_{22 eff}=V_{22}(x)-V_{12}(x)$.

All this leads, e.g., to two branches of potential oscillations
accompanied by natural effect for the band structure (widening or
narrowing of forbidden zones).

2. Let the thresholds be unequal  $\varepsilon_{1} <
\varepsilon_{2}$,
 $V_{11}(x)=V_{22}(x)=0$ and the coupling of channels be some constant
 $V_{12}(x)=V_{21}(x)= const$.  Let us choose the boundary
conditions as follows: $\Psi_{i}(0)=\Psi_{i}(\pi)=0;\,i=1,2 $. It
means the requirement of equidistancy of knots in different
channels with quite different partial channel energies
$E_{i}=E-\varepsilon_{i} $. This 'paradoxical' situation can exist
 despite the difference of the thresholds if the
effective kinetic energies in both the channels, e.g., ground
state are equal.  This can be achieved by choosing different
spectral weights of partial components which are proportional to
one another:  $\Psi_{1}(x)= c_{1}/c_{2} \Psi_{2}(x)$. Really, the
partial channel equations can be rewritten as uncoupled ones using
the substitution $\Psi_{1}(x)$ by $ c_{1}/c_{2} \Psi_{2}(x)$ and
vice versa:  \begin{eqnarray} -\Psi_{1}''(x)= (E-\varepsilon_{1} -
V_{12} c_{2}/c_{1}) \Psi_{1} -\Psi_{2}''(x)= (E-\varepsilon_{2} -
V_{21} c_{1}/c_{2}) \Psi_{2}.  \end{eqnarray} The expressions in
brackets in the right hand sides are effective kinetic energies
and they become equal if
$$c_{2}/c_{1}= \Delta \varepsilon _{12}/2 V_{12} \pm \sqrt{\Delta
\varepsilon_{12}^{2}/(4 V_{12}^{2})+1 }.$$

3. Again let  $\varepsilon_{1} <  \varepsilon_{2}$,
$V_{11}(x)=V_{22}(x)=0$, but with the modified boundary conditions
 $\Psi_{i}(- \pi/i=\Psi_{i}( \pi/i)=0;\,i=1,2 $   and interchannel
coupling $V_{12}(x)=V_{12} \delta (x)$. In this model the
effective kinetic energy in the second channel with {\it smaller}
partial energy $E-\varepsilon _{2}$ must be  even {\it greater}
than in the first channel in order to provide the {\it two times
smaller distance between the knots} ($\pi $ in comparison with $2
\pi$ for the first channel).

\newpage
Figure captions

Fig 1. The potential perturbation of the bottom of the infinite
rectangular well which results in lowering  (shown by arrow
$\Delta E_{1}$) only one level $\stackrel{\circ}{E}_{1}=1 \to
E_{1}=-4$ of the ground state.  The bold black arrow directed
downward points to the well (dashed-dotted  line $\Delta V(x)$)
acting on the most sensitive, central, region of the wave function
to shift the level downward. The same arrows directed upward near
the walls of the initial well, where the function $\Psi_{1}$ has
knots and where it is the least sensitive to potential changes,
point to the barriers needed to keep all other energy levels $E_{i
\ne 1}$ from shifting. The transformations of the wave functions
$\stackrel{\circ}{\Psi}_{1,2} \to \Psi_{1,2}$ are shown by thin
dashed lines.  The instructive value of this picture consists in
that it demonstrates qualitative features of {\it universal,
elementary } transformation for a {\it single} bump of a wave
function. It allows one to understand the rule of the wave
transformations of arbitrary states with {\it many} bumps and in
{\it arbitrary} potentials, see also other pictures.  b) Evolution
of the $\Psi_{1}$ when $\Delta E = -1$, $-3$. Note that the knots
at the edges of the interval (equal to period, see text) do not
move c) Evolution of the potential when $\Delta E = -1$, $-2$,
$-3$. Do not confuse the cases b,\,c) with motions in the system:
we mean changes of $\Psi_{1}$, $V$ under successive transitions
from one system to another ("in the space of different models").

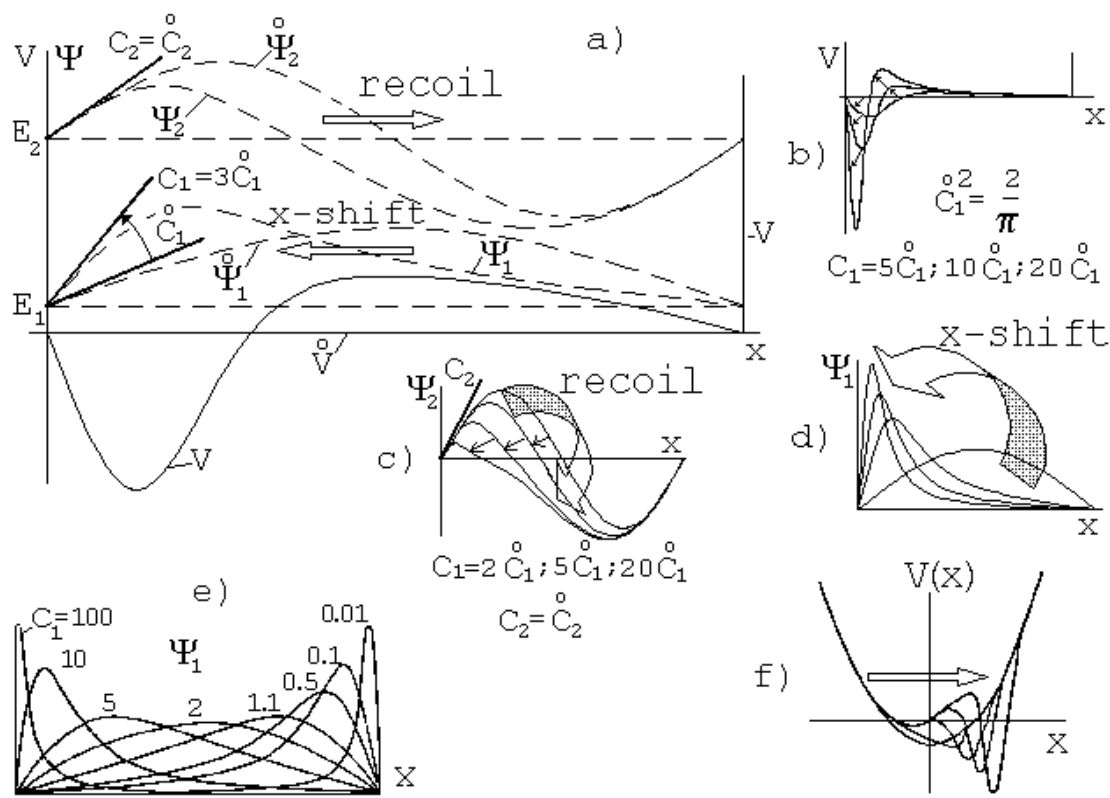
Fig.2.  The (a,b) transformation  of the infinite rectangular
potential well $\stackrel{\circ}{V}\rightarrow V$ and
 eigenfunctions $\stackrel{\circ}{\Psi}_{1}(x)$ (a,d),
 $\stackrel{\circ}{\Psi}_{2}(x)$ (a,c) by increasing spectral weight
factor (SWF) $\stackrel{\circ}{c}_{1}\rightarrow c_{1}$, the
derivative
 $\Psi'_{1}(x=0)$ at the left wall.  The scales of the functions
$\Psi_{1},\,\Psi_{2}$  (a) are shifted up to the corresponding
energy levels  $E_{1},\,E_{2}$.  In (b,c,d) the evolution of the
potential and functions with increasing $c_{1}$ 2, 5, 10, 20 times
is demonstrated.
 Meanwhile {\bf all !} energy levels $E_{n}$ and all SWF, except  one $c_{n
 \neq 1}$, remain unchanged, as is seen for  $\Psi_{2}$ (a,c). The
parameter SWF $c_{1}$ controls the localization of the wave
function $\Psi_{1}(x)$ (a,d,e) in space: by increasing  $c_{1}$
the ground state is pressed to the left potential wall, as is
shown by arrows in (a,d). This is performed by the potential
barrier in (a,b) on the right which shifts the function
$\Psi_{1}(x)$ to the left, and the potential well on the left
which simultaneously  compensates the influence of the barrier on
{\it the energy levels}  and keeps them all at the same places.
All wave functions, except the ground state, undergo some recoil
in the opposite direction which is demonstrated by $\Psi_{2}(x)$
(a,c). So there is separation of the bound state from others.

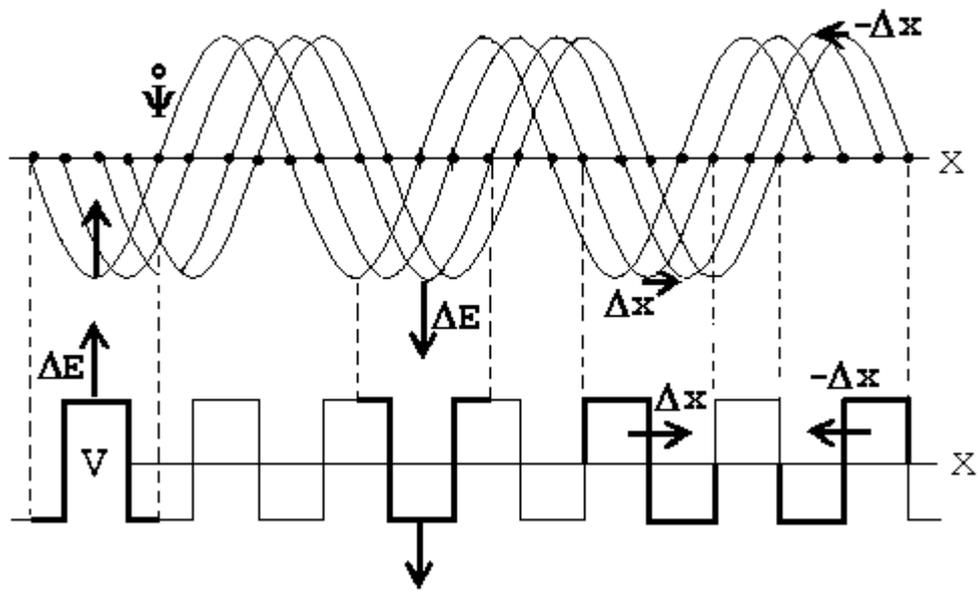
Fig.3. The wave functions of a free motion $\sin(kx + \alpha )$ at
energy $E=k^2$ corresponding to frequency  of periodic potential
$V(x)$ oscillations. These solutions of the Shr\"odinger equation
are chosen  with different values $\alpha $ so that the knots
coincide with: the middle of the potential barriers, or wells, or
left sides of the barriers (wells). All free waves have the
equidistant distribution of knots, and switching on the perturbing
periodic potential must violate this. However, we can find such
values of $E_{>}; E_{<};$ etc. (shifted from $E_{1}$, see vertical
arrows) at which we can choose perturbed solutions with the same
equidistant distributions of knots. Horizontal arrows correspond
to the shifts of space localization of the wave bumps (see
Fig.4.C,D) without shifting the knots.

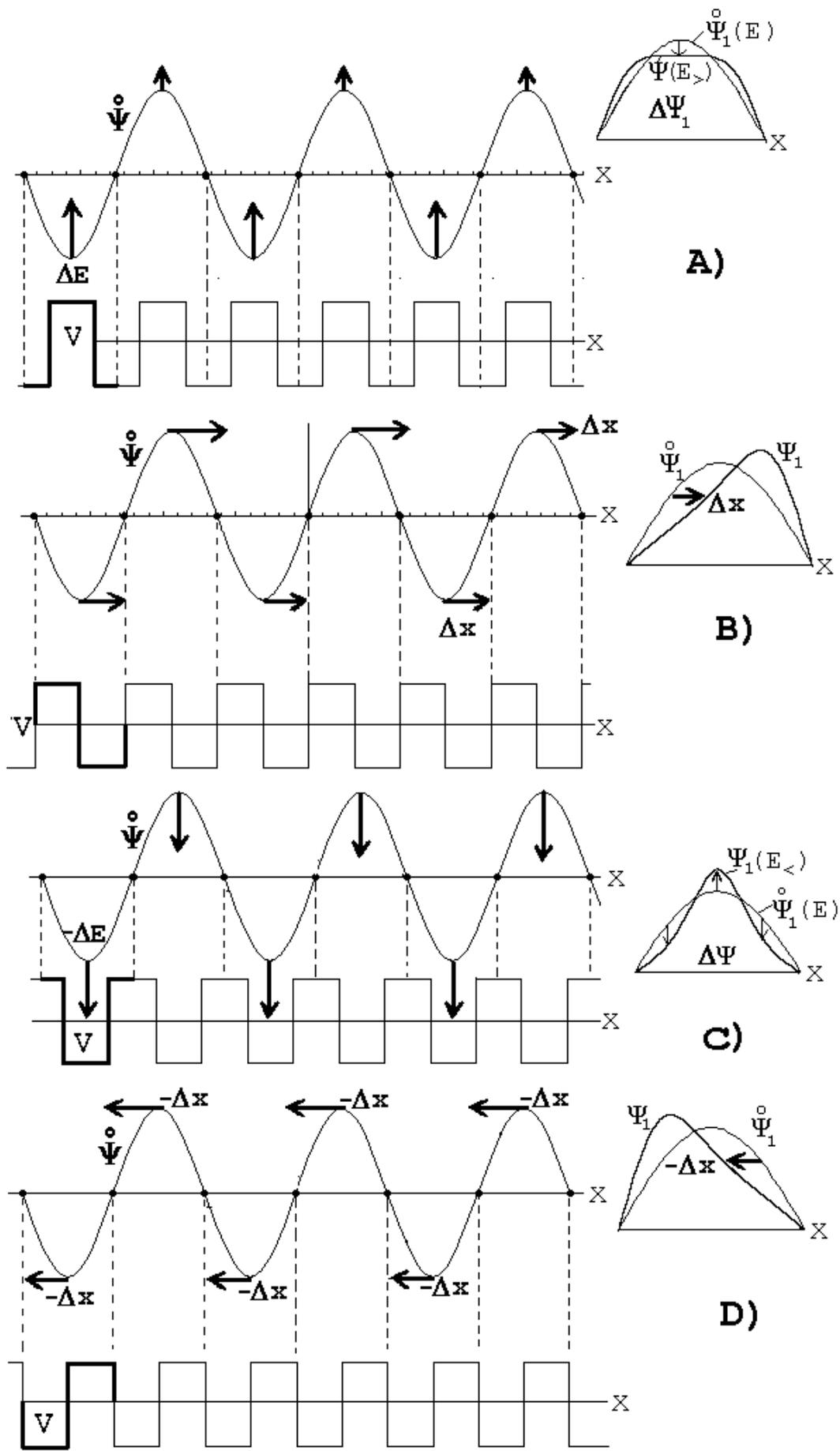
Fig.4.  The same as in Fig.3, but the cases with a different shift
of knots are considered separately: A,B) the perturbed solution is
shifted up without violation of the distance between knots; B)
shift to the right (mainly) of localization of any wave bump; C)
shifting down the solution with keeping equidistant distribution
of its knots.
 D) shift
to the left.  Transformations of wave bumps are shown separately
at the left sides of the figures (see also Fig.5).

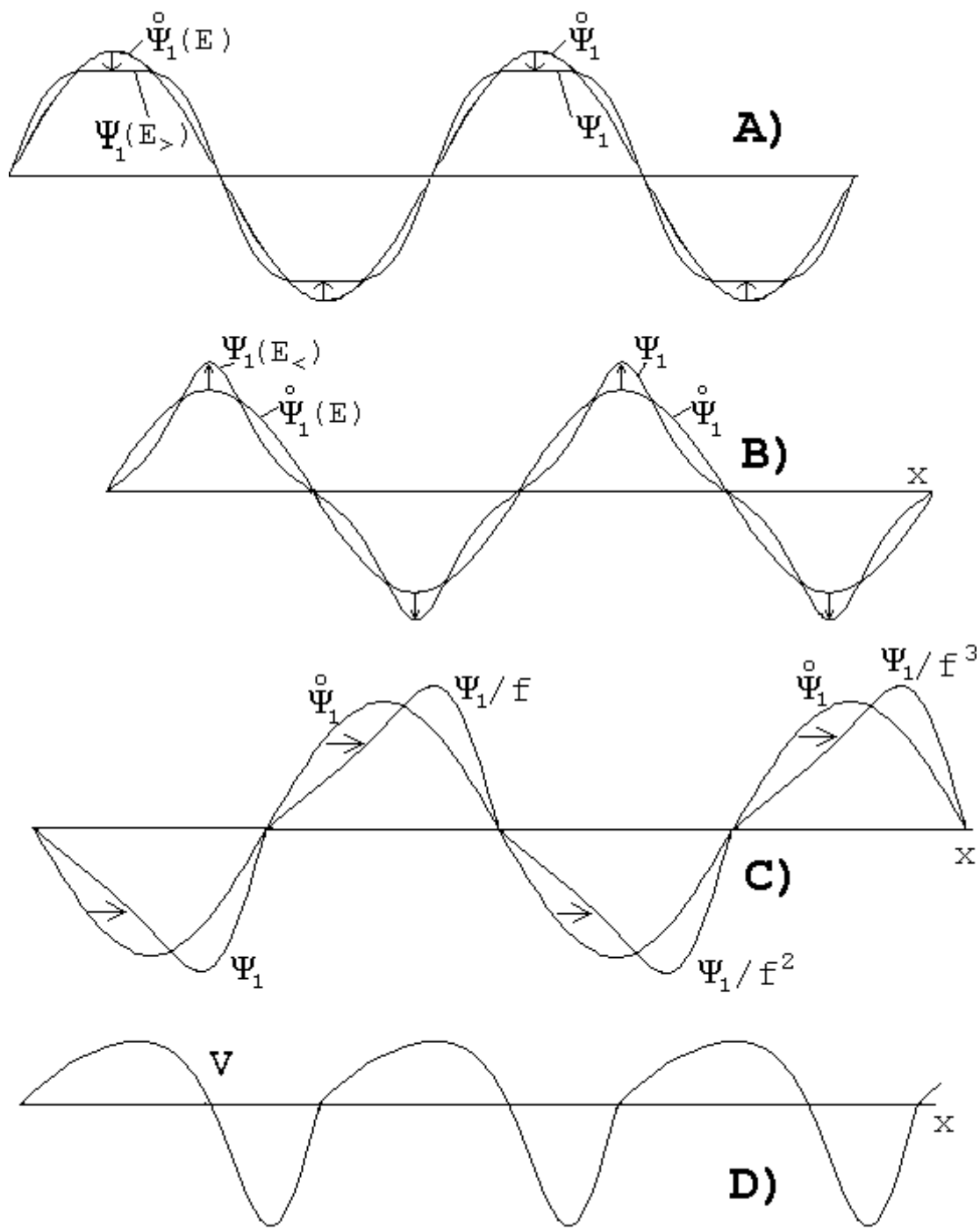
Fig.5.  Perturbations of the initial free wave functions with the
different x-shifts ($\alpha $) by the periodic potential with the
period equal to the distance between the wave knots. The perturbed
solutions are considered at energy values (A:\,$E_{>};\,B:\,
E_{<};\, $, etc.) where their knots do not change their positions.
C) Exponential increase of swinging amplitude of solutions in
forbidden zone is demonstrated by introducing big negative factors
$f,f^2,f^3...$ which provide the smooth junction of solutions on
neighbor periods (the corresponding potential perturbation is
shown in Fig.5.D.).

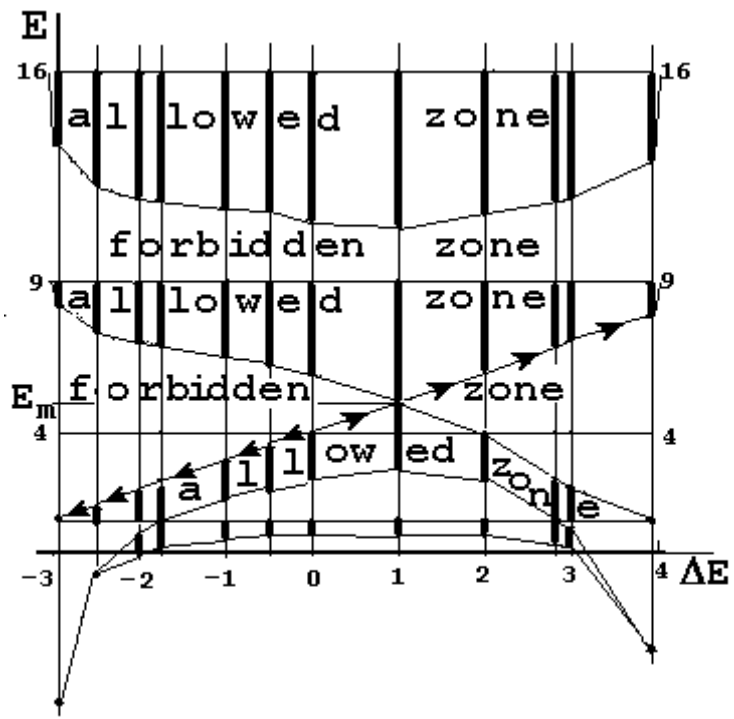
Fig.6. Shifting the upper boundary of the second allowed spectral
zone (shown with arrows). Pay attention to the merging of this
zone with the upper one at $\Delta E=1$. Find two another examples
of merging of first and second zones.

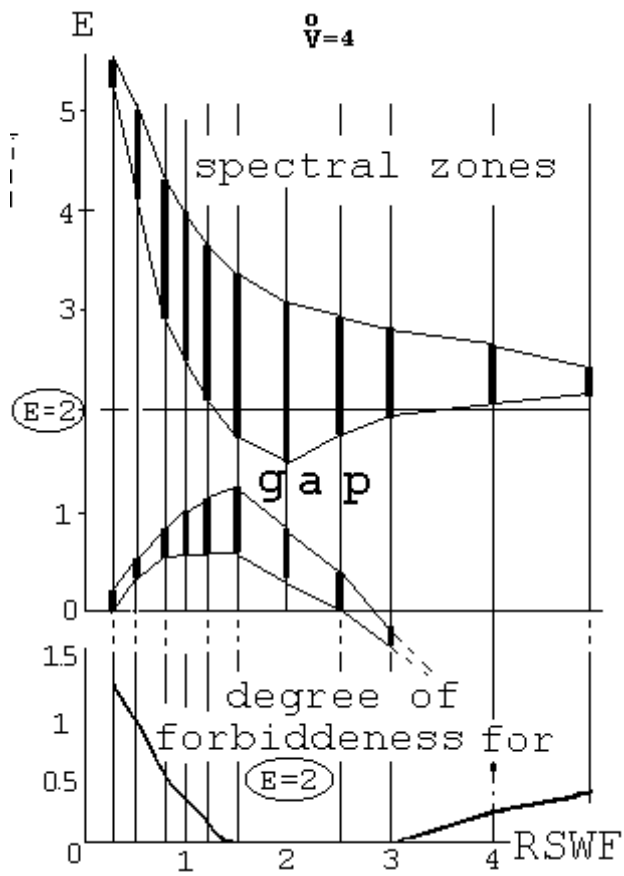
Fig.7. (Upper part)  Changes in zone structure corresponding to
relative variations of a derivative at one end of a period
(spectral weight factor) $c/\stackrel{\circ}{c}$ at energy point
$E=2$.  The initial periodic potential $\stackrel{\circ}{V} \delta
(x-n \pi )$ has Dirac comb peaks with the strength
$\stackrel{\circ}{V}=4$ and period $\pi $. The wave function at
$E=2$  is on each period an eigenfunction of the eigenvalue
problem with the boundary conditions on the edges of a period
specified so that the auxiliary discrete energy level just
coincides with the chosen point $E=2$.  (Lower part) Changes of
the imaginary part of quasi-momentum $Im K(E)$ which characterize
the degree of forbiddeness, the index  of exponential swinging of
solutions.

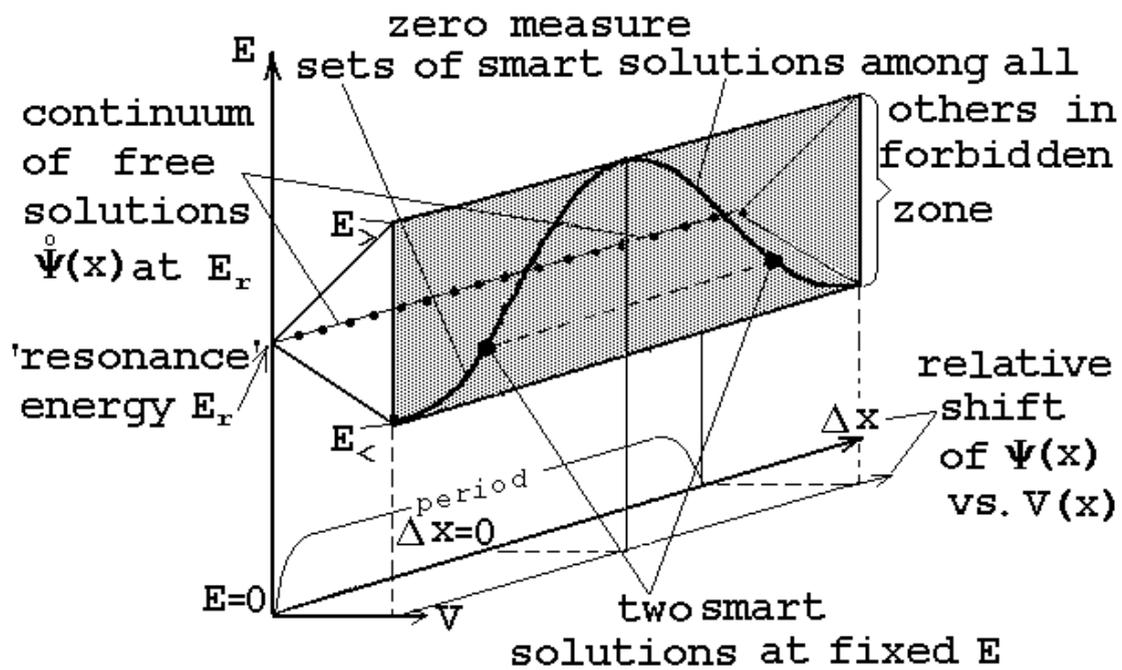
Fig.9.  Schematic illustration of the set of 'smart' solutions
with periodic knots  (curved line)  which are a zero measure
manifold among all other solutions in the chosen forbidden zone
(rectangular shadowed area corresponding to $\pm \Delta x$ on the
finite interval having  a length of the period).

 The continuum of free periodic solutions of the type shown in  Fig.3
  at 'resonance' energy $E_{r}$
(shown here as a line with points) are transformed by periodic
perturbation
 $V$ into exponentially increasing (decreasing) solutions of a forbidden
zone which splits the initial continuous spectrum.

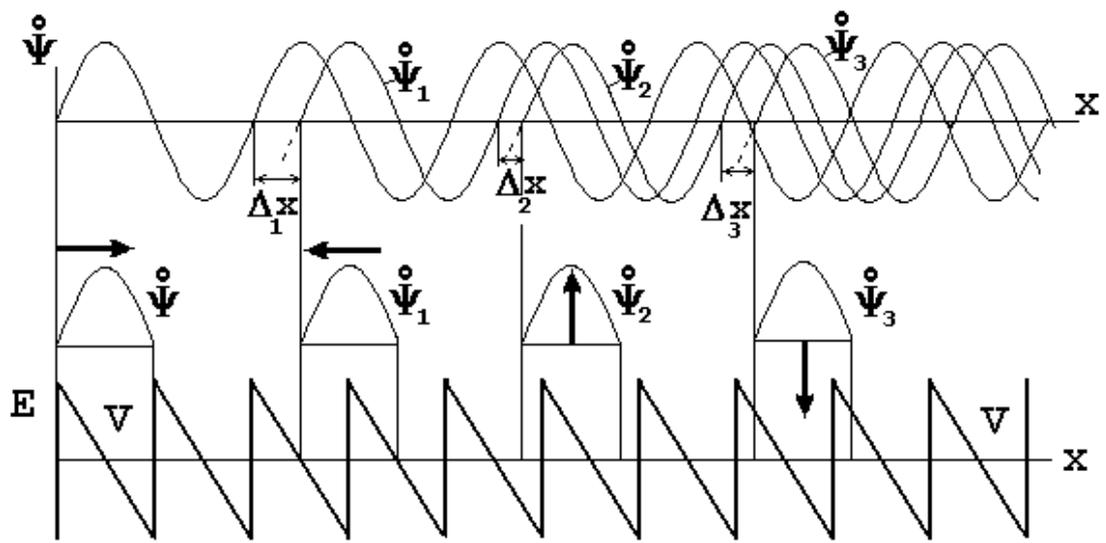
\end{document}